# Diurnal astronomy: using sticks and threads to find our latitude on Earth

**Néstor Camino**
Complejo Plaza del Cielo. Physics Dep., Facultad de Ingeniería, Universidad Nacional de la Patagonia San Juan Bosco, Esquel, Argentina.

**Alejandro Gangui**
IAFE/Conicet and Universidad de Buenos Aires, Argentina.

*It is well known that the length and orientation of a shadow cast by a vertical gnomon depends on the time of the day and on the season of the year. But it also depends on the latitude of the site of observation. During the equinoxes, the temporal sequence of the shadows cast by each of the points that form any object follows a straight line from West to East. A simple construction using sticks and threads can be used to materialize the plane of Celestial Equator's local projection, giving us a way to calculate our astronomical latitude during daytime with high precision.*

**The movement of the Sun in the local sky**

There is nothing surprising for our high-school students (and for most people) when we tell them that the Earth rotates on its axis and that this movement is responsible for the day-and-night cycle. It *might* be a surprise to them, however, when we insist on using an observer-based (*topocentric*) model to study the Sun's "movement" during the day, and how this apparent movement changes during the year. Diurnal astronomy is in fact much easier to learn, as a first step, when we describe all phenomena from the observer's reference frame.

Daily, the Sun follows an arc on the sky, starting off on the Eastern horizon (the rising Sun region) and ending on the Western horizon (where the Sun sets). This arc is not arbitrary, for its bent is related to the latitude of the observer. For people living in the northern hemisphere, this arc always leans against the southern direction, with its summit (corresponding to solar noon) pointing exactly towards the South and with its western and eastern legs being symmetric with respect to the North-South horizontal line. Something similar happens for people living in the southern hemisphere, but mirror-reflected (see Figure 1). By "continuity", people living exactly on the Equator will see the arc of the Sun with both legs perfectly perpendicular to both (rising and setting) horizons.

We know this arc followed by the Sun in the local sky is the resulting (visible) effect of our own (Earth's) rotation. Hence, the plane containing this arc must be always parallel to the Equator. The degree of inclination of this fundamental plane (and of the diurnal arc) with respect to the vertical (the zenith-observer line) yields the local latitude. For example, the bent of the plane is null for people living on the Equator (the plane is parallel to the vertical direction); hence, their latitude is zero. For those living on the Poles, the arc of the Sun (and the plane containing it) is parallel to the horizon, and the corresponding inclination is 90 degrees.

**The movement of the Sun in the local sky during an Equinox**

Two very special days in the year are the Equinoxes (of March and of September). During these days, regardless of the position of the observer on the planet, every earthling will see the Sun traveling exactly twelve hours along its diurnal arc, rising exactly on the East and setting exactly on the West. And it is precisely on these days that the many shadows cast by the tip of a gnomon during the whole day will be aligned into a straight line going from West to East.

Now, let us mark (with small sticks, for instance) the points on the floor where the tip of the gnomon projected its shadow (say, one for every half-hour, or may be less). Joining these sticks with the tip of the gnomon using threads, we will find that the collection of threads forms a sector of a plane (see Figure 2). As expected, this plane (materialized with threads) will be qualitatively similar for every location on the Earth. However, the different bents of the arc of the Sun for different latitudes will result in different bents for these planes (that we call the *Celestial Equator's local projections*).

Once we have the plane thus constructed, measuring just one angle, the one formed between the vertical gnomon and the threads' plane, we get the latitude of our site. Moreover, the direction orthogonal to this plane points towards the celestial pole (say, towards Polaris, approximately, if we are doing the demonstration in the northern hemisphere). And one can do this during the day, when stars are not visible. Finally, the line on the floor orthogonal to the East-West line formed by the shadows of the tip of the gnomon yields the *meridian line* of the observing site, namely, the astronomical or geographical local North-South direction. This direction does not coincide, in general, with what one gets by employing a magnetic needle or a standard compass, which point towards the magnetic North direction.

On March 20, 2009, the day of the southern autumnal Equinox, we carried out this outdoor activity during the whole day in five different locations of South America. Figure 3 shows schematically what we obtained in each place. The activity emphasized aspects that were characteristic to each location (like times of solar noon, length of the shadows, inclination of equatorial planes) and others that were common to all cities, such as the straight line followed by the tip of the shadows on that date (which materialized the local East-West line), the length of day and night, the points of sunrise and sunset on the local horizon, etc.

We worked together with physics and natural science secondary school teachers and with their university professors, as well as with pre-service primary and secondary school teachers. Nine hours under the Sun allowed us to discuss a whole range of topics related to both physics and astronomy, and how to implement the new knowledge in our classrooms.

There is still a lot to do to appropriately reinstall astronomy in both primary and secondary school in our countries. We think that activities like the one described here, which articulate experiences from the university and from school, are of great benefit to us all, and we plan to intensify our joint work in the future.

**Acknowledgements**

We would like to thank our students and collaborators from Esquel and Buenos Aires, and our colleagues from Argentina, Brazil and Uruguay, members of the CTS Project "Education of Astronomy", for helping us prepare and carry out the Equinox activities described in this paper. A.G. acknowledges support from Conicet and from the University of Buenos Aires.

Néstor Camino *is researcher in the teaching of astronomy and is professor of physics at the National University of Patagonia San Juan Bosco and at the Instituto Superior de Formación Docente N°804 in Esquel.*

Alejandro Gangui *is staff researcher at the Institute for Astronomy and Space Physics (IAFE) and is professor of physics at the University of Buenos Aires (UBA).*

# Figures

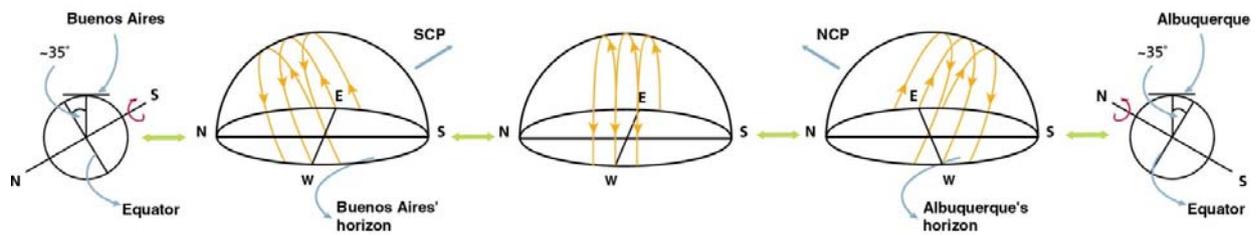

Figure 1: The diagram shows the smooth transition between an *external* representation for the day-and-night cycle (sketches on the far left and far right) and an *internal*, observer-based, one (the three sketches of the middle). The orange curves are the three solar arcs corresponding to the four special moments (days) of the Earth's orbit around the Sun: two Solstices (longest and shortest arcs) and two Equinoxes (both represented by the arc in the middle, the only times the Sun rises exactly on the East and sets exactly on the West). If one is located in the southern hemisphere, in Buenos Aires for example, the South celestial pole (SCP) is always above the horizon (see the two sketches on the left). Analogously, for people living on the northern hemisphere, in Albuquerque, NM, for example, the North celestial pole (NCP) lies always above their horizon (see the two sketches on the right). The sketch in the middle corresponds to people living exactly on the equator of the Earth. Travelling along a meridian one can go from northern latitudes (on the right) to southern latitudes (on the left). The smooth transition of the solar arcs in the sketches in the middle of the Figure shows what this observer would see as she travels southwards: initially leaning against the southern direction, the arcs of the Sun continuously straighten up, and finally bend towards the northern direction, when the traveler sets foot in the southern hemisphere.

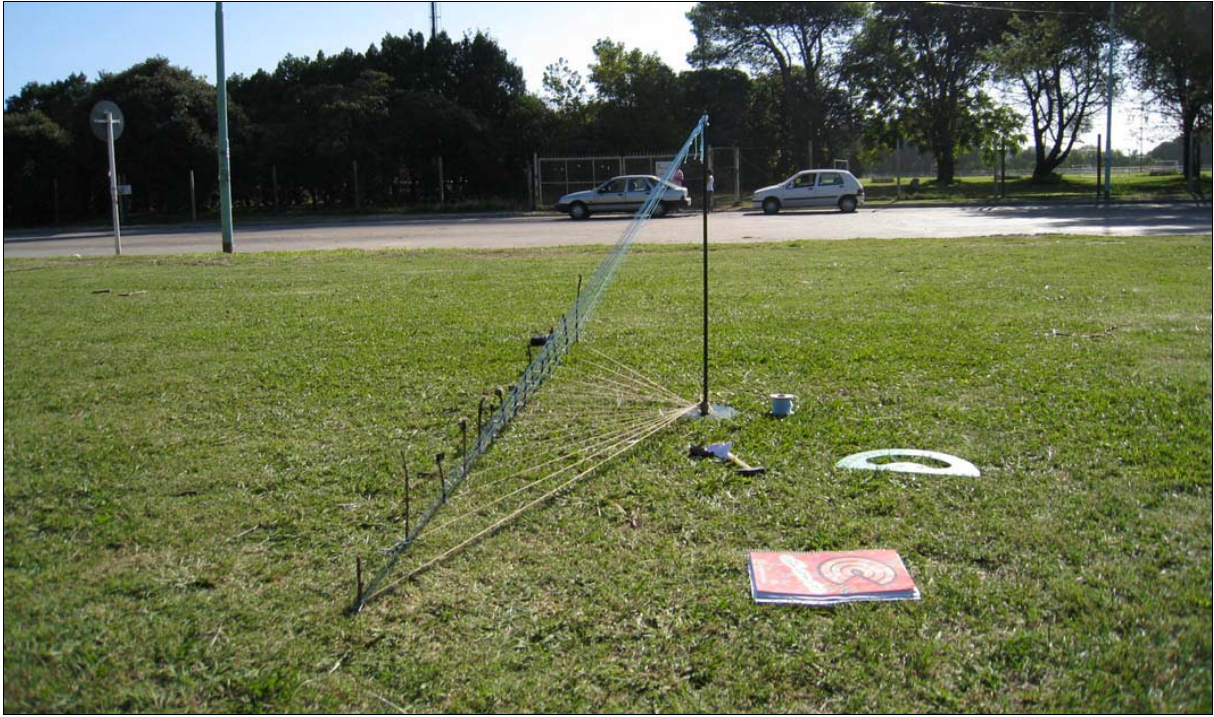

Figure 2: Plane formed by the collection of the threads (that join the marks left on the floor by the shadows of the vertical gnomon) with the tip of the gnomon. The angle between the gnomon and the plane gives the latitude of the site, in this case, of the city of Buenos Aires, at a latitude of approximately 35 degrees South. Correspondingly, astronomical North is to the right of the image, while the South celestial pole is located in a direction orthogonal to the plane, towards the upper-left part of the image.

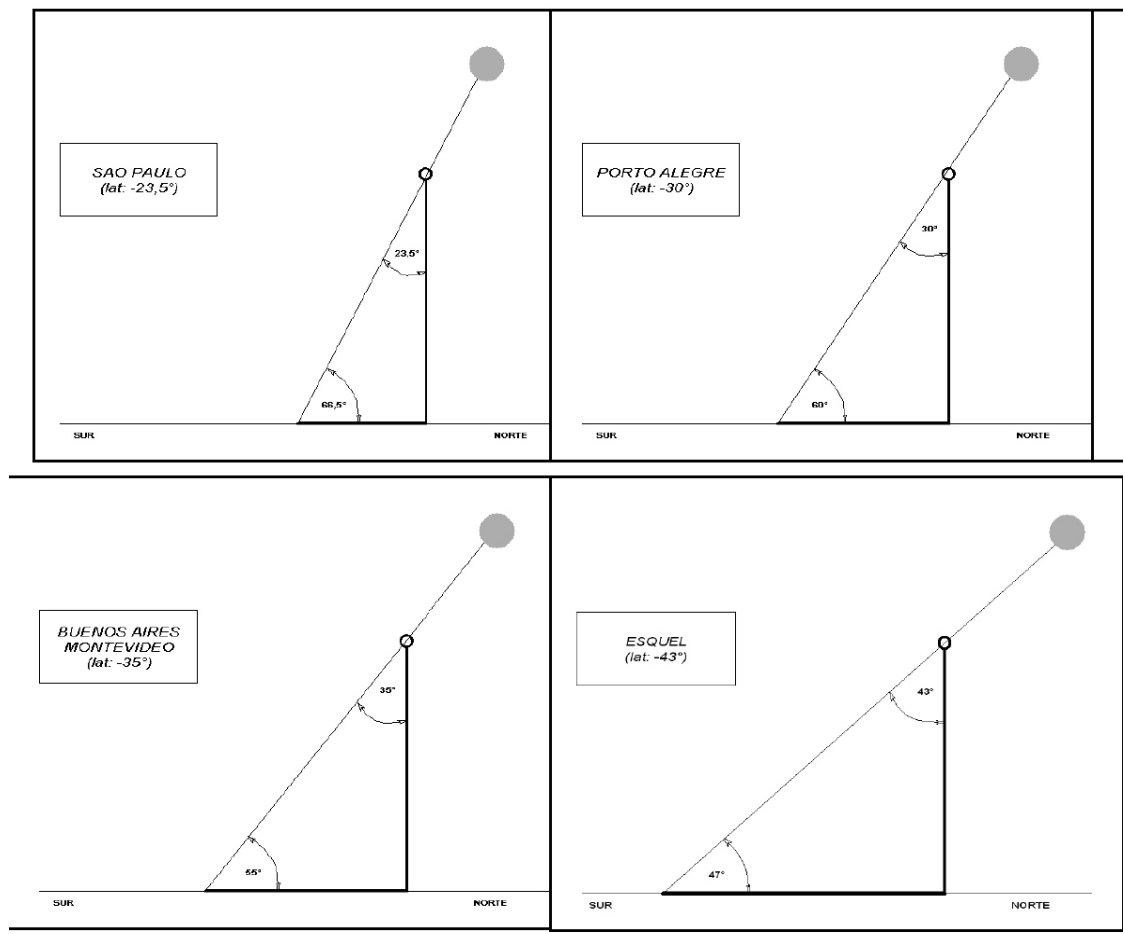

Figure 3: The local meridian planes, at true solar noon, showing the different angles measured in different locations, from São Paulo in Brazil to Esquel in Patagonia, Argentina. In each drawing, the Sun is depicted on top and to the right of each box, and the diagonal lines (Sun rays) emerging from it indicate the Celestial Equator's projection plane corresponding to each of the five latitudes, as seen from the side (cf. Figure 2).